\documentclass[11pt, letterpaper]{article}
\usepackage[utf8]{inputenc}
\usepackage{amsmath,amssymb}
\usepackage{graphicx}
\usepackage{subfig}
\usepackage{here}
\usepackage{float}
\usepackage{multirow}
\usepackage[margin=0.75in]{geometry}
\usepackage{color}
\usepackage{ulem}
\usepackage{authblk}
\usepackage{siunitx}
\usepackage{booktabs}
\DeclareMathAlphabet{\mathpzc}{OT1}{pzc}{m}{it}
\newcommand{\vev}[1]{\langle\Omega|#1|\Omega\rangle}

\newcommand{\lsr}{\mathcal{R}_{0}}
\newcommand{\hypgeom}[2]{{}_{#1}F_{#2}}
\newcommand{\double}[2]{(#1,\,#2)}
\newcommand{\dif}[1]{d #1}
\newcommand{\glueFourD}{\big\langle \alpha G^{2} \big\rangle}
\newcommand{\glueSixD}{\big\langle g^{3} G^{3} \big\rangle}

\newcommand{\rom}[1]{\uppercase\expandafter{\romannumeral #1\relax}}
\title{A QCD Sum-Rules Analysis of Vector ($1^{--}$)  Heavy Quarkonium Meson-Hybrid Mixing}
\author[1]{A.\ Palameta}
\author[1]{J.\ Ho}
\author[2]{D.\ Harnett}
\author[1]{T.G.\ Steele}
\affil[1]{Department of Physics and Engineering Physics\\ University of Saskatchewan\\           Saskatoon, SK, S7N 5E2, Canada}
\affil[2]{Department of Physics\\ University of the Fraser Valley\\ Abbotsford, BC, V2S 7M8, Canada}
\begin{document}
\maketitle
%%%%%%%%%%%%%%%% Abstract %%%%%%%%%%%%%%%%
\begin{abstract}
\noindent We use QCD Laplace sum-rules to study meson-hybrid mixing in vector ($1^{--}$) heavy quarkonium. We compute the QCD cross-correlator between a heavy meson current and a heavy hybrid current within the operator product expansion. In addition to leading-order perturbation theory, we include four- and six-dimensional gluon condensate contributions as well as a six-dimensional quark condensate contribution. We construct several single and multi-resonance models that take known hadron masses as inputs.  We investigate which resonances couple to both currents and so exhibit meson-hybrid mixing. Compared to single resonance models that include only the ground state, we find that models that also include excited states lead to significantly improved agreement between QCD and experiment. In the charmonium sector, we find that meson-hybrid mixing is consistent with a two-resonance model consisting of the $J/\psi$ and a 4.3~GeV resonance. In the bottomonium sector, we find evidence for meson-hybrid mixing in the $\Upsilon(1S)$, $\Upsilon(2S)$, $\Upsilon(3S)$, and $\Upsilon(4S)$. 
\end{abstract}

%%%%%%%%%%%%%% Introduction %%%%%%%%%%%%%%
\section{Introduction}\label{I}
Hybrids are hadrons containing explicit gluon degrees of freedom in addition 
to a constituent quark and antiquark.
They are colour singlets and so should be allowed within QCD.
However, they have yet to be conclusively identified in experiment
(see, e.g., ref.~\cite{Meyer:2015eta} for a comprehensive review).

Hybrids can be broadly classified as having quantum numbers (i.e., $J^{PC}$) 
that are exotic or non-exotic.
Exotic quantum numbers (e.g., $0^{--},\ 0^{+-},\ 1^{-+},\ 2^{+-}$) are those 
not accessible to conventional quark-antiquark ($q\overline{q}$) mesons;
the rest of the quantum numbers are non-exotic and are accessible to both $q\overline{q}$-mesons and hybrids.
Looking for resonances with an exotic $J^{PC}$ is a promising hybrid search strategy
being used, for example, at GlueX.
Furthermore, hybrids with exotic $J^{PC}$ would be unable to quantum mechanically
mix with $q\overline{q}$-mesons (as no conventional meson could have the $J^{PC}$ in question),
and so could perhaps appear as pure, unmixed states.
In contrast, hybrids with non-exotic $J^{PC}$ are expected to mix with $q\overline{q}$-mesons
resulting in hadrons that would be superpositions of both conventional meson and hybrid.

In this article, we consider meson-hybrid mixing in (non-exotic) vector 
($1^{--}$) charmonium ($c\overline{c}$) and bottomonium ($b\overline{b}$).
The heavy quarkonium sectors have received considerable attention lately due 
primarily to the discovery of the XYZ resonances
(see~\cite{Brambilla:2010cs,Eidelman:2012vu} for reviews
and~\cite{BESIII:2016adj} for some recent developments).
These XYZ resonances are a collection of hadrons 
many of whose properties (e.g., masses, widths, and decay rates) do not agree
with quark model predictions~\cite{BarnesCloseSwanson1995}.
Unsurprisingly, the XYZ resonances have generated a lot of discussion concerning 
outside-the-quark-model hadrons such as hybrids.
We focus on $1^{--}$ rather than some other $J^{PC}$ because more is known 
about the spectra of $1^{--}$ heavy quarkonium than is known about the spectra 
for the other quantum numbers~\cite{Olive:2016xmw}.

We investigate meson-hybrid mixing with QCD Laplace sum-rules 
(LSRs)~\cite{Shifman:1978bx,Shifman:1978by,Reinders:1984sr,narisonbook:2004}.
Using the operator product expansion (OPE)~\cite{Wilson:1969zs}, 
we compute the cross-correlator between a $q\overline{q}$-meson current and a hybrid current
(see (\ref{CurMes}) and~(\ref{CurHyb}) respectively below).
In the cross-correlator calculation, we include leading-order (LO) QCD contributions from perturbation theory and non-perturbative 
corrections due to the four-dimensional (4d) and 6d gluon condensates as well as the 6d quark condensate.
We then analyze several single and multi-resonance models of the hadron mass spectra that take known 
resonance masses as inputs. We determine which resonances couple to 
both currents and so can be considered mixed.
The QCD sum-rules methodology has been applied to hadron mixing problems in
a number of systems including
pseudoscalar meson-glueball mixing~\cite{Narison:1984bv},
scalar meson-glueball mixing~\cite{Harnett:2008cw},
$1^{++}$ charmonium hybrid-$\overline{D}D^{*}$ molecule mixing~\cite{Chen:2013pya}, 
and open-flavour heavy-light meson-hybrid mixing~\cite{ho:2017}.

We find that multi-resonance models that include excited states in addition to 
the ground state lead to significantly improved agreement between QCD and experiment 
when compared to single resonance models that include only the ground state. 
In addition, we show explicitly that the higher mass excited states make numerically significant contributions to the LSRs despite the tendency of LSRs to suppress such resonances.
Finally, we find that meson-hybrid mixing in the charmonium sector is described well 
by a two-resonance model consisting of the
$J/\psi$ and a 4.3~GeV state such as the X(4260).
In the bottomonium sector, we find evidence for meson-hybrid mixing in all of the
$\Upsilon(1S)$, $\Upsilon(2S)$, $\Upsilon(3S)$, and $\Upsilon(4S)$.

%%%%%%%%%%%%% The Correlator %%%%%%%%%%%%%
\section{The Correlator}\label{II}
We consider the following cross-correlator
\begin{align}
  \Pi_{\mu\nu}(q) &= i\!\int d^{4}\!x \;e^{iq\cdot x} 
    \vev{\tau\, j_{\mu}^{(\text{m})}(x)\; j_{\nu}^{(\text{h})}(0)}
    \label{CorFn}\\
   &=\left(\frac{q_{\mu}q_{\nu}}{q^2}-g_{\mu\nu}\right)\Pi(q^2)
   \label{CorFnProj}
\end{align}
between quarkonium meson current
\begin{equation}\label{CurMes}
  j_{\mu}^{(\mathrm{m})} = \overline{Q}\gamma_{\mu} Q
\end{equation}
and quarkonium hybrid current~\cite{GovaertsReindersWeyers1985}
\begin{equation}\label{CurHyb}
  j_{\nu}^{(\mathrm{h})}=  \frac{g_{s}}{2}\overline{Q}\gamma^{\rho}\gamma^{5}\lambda^{a} \widetilde{G}^{a}_{\nu\rho} Q
\end{equation}
where
\begin{equation}\label{dualFieldStrength}
  \widetilde{G}^{a}_{\nu\rho} = \frac{1}{2}\epsilon_{\nu\rho\omega\zeta}G^{a}_{\omega\zeta}
\end{equation}
is the dual gluon field strength tensor and $\epsilon_{\nu\rho\omega\zeta}$ is the 
totally antisymmetric Levi-Civita symbol.
The function $\Pi$ in~(\ref{CorFnProj}) probes $1^{--}$ states.

\begin{figure}
\centering
\begin{tabular}{ccc}
\includegraphics[width=50mm]{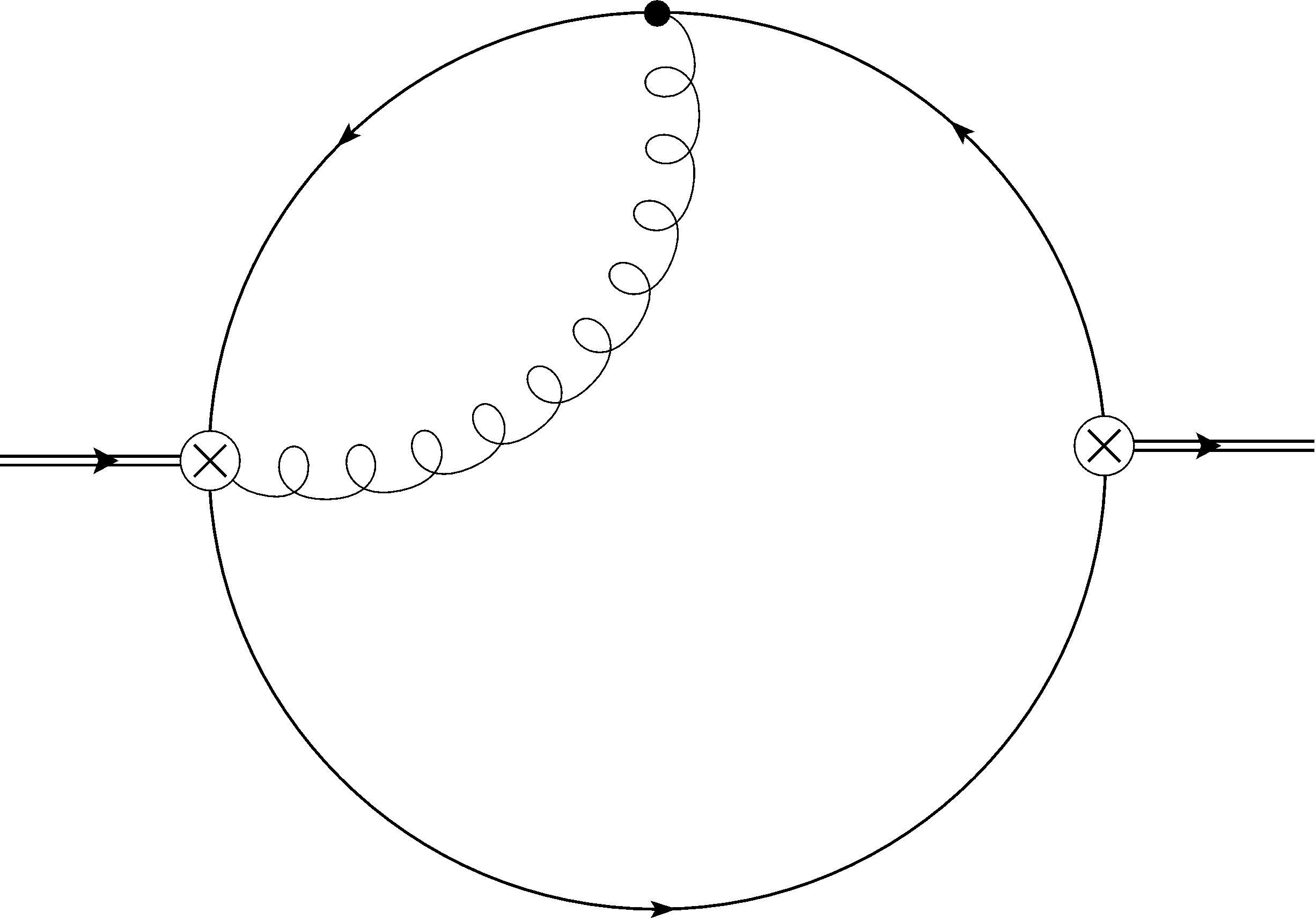} & \includegraphics[width=50mm]{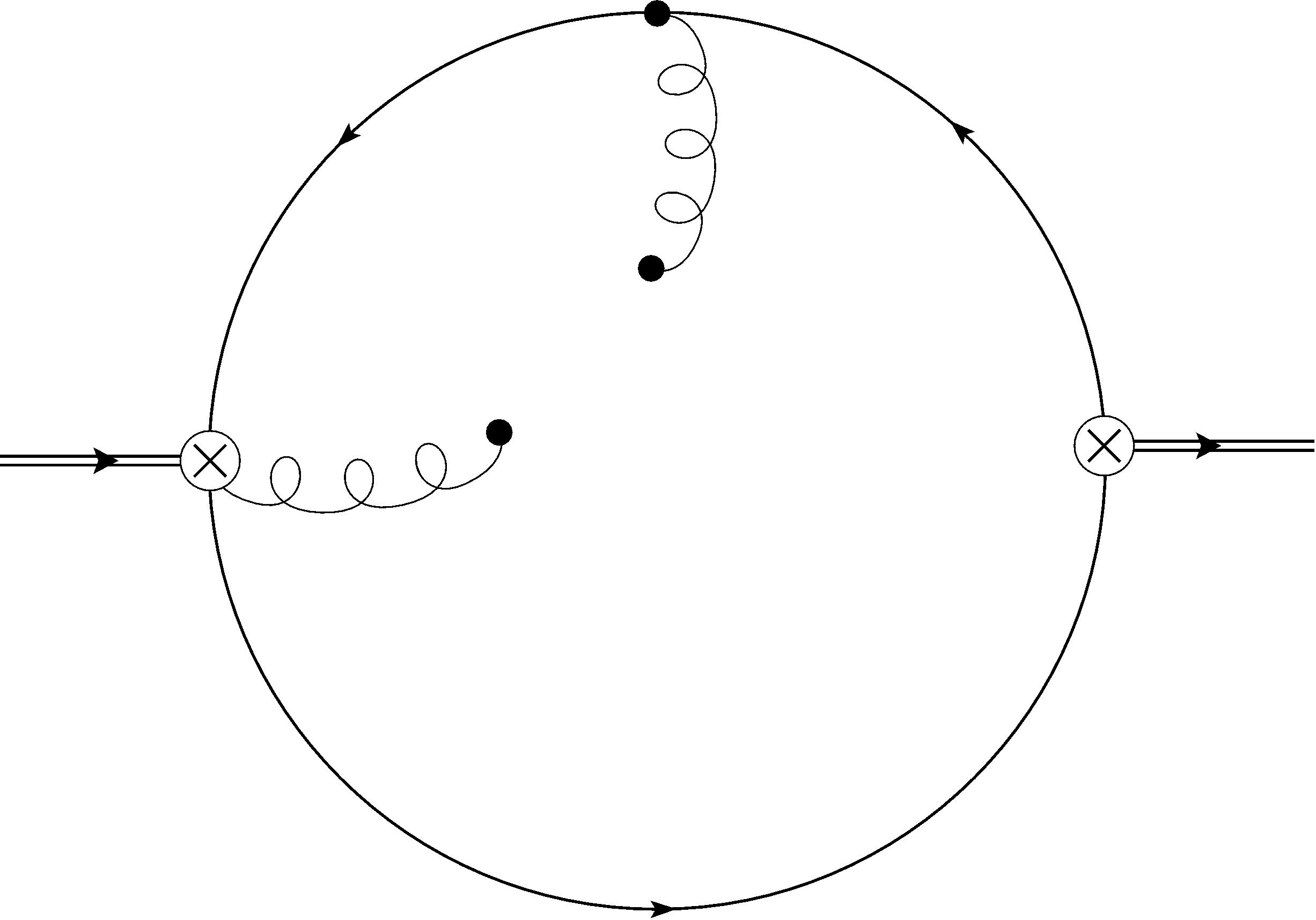} & \includegraphics[width=50mm]{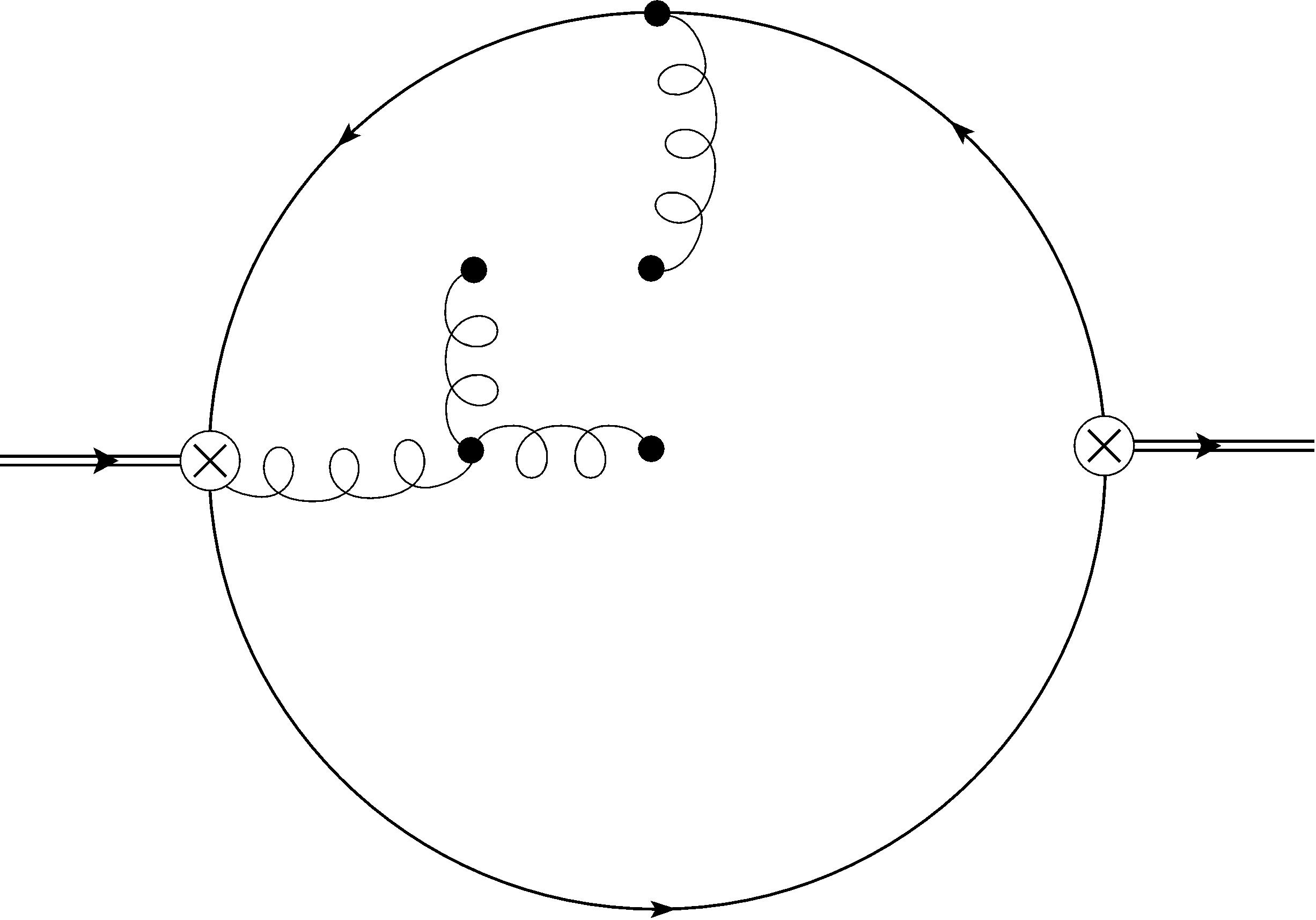}\\
Diagram \rom{1} & Diagram \rom{2} & Diagram \rom{3} \\[15pt]
\includegraphics[width=50mm]{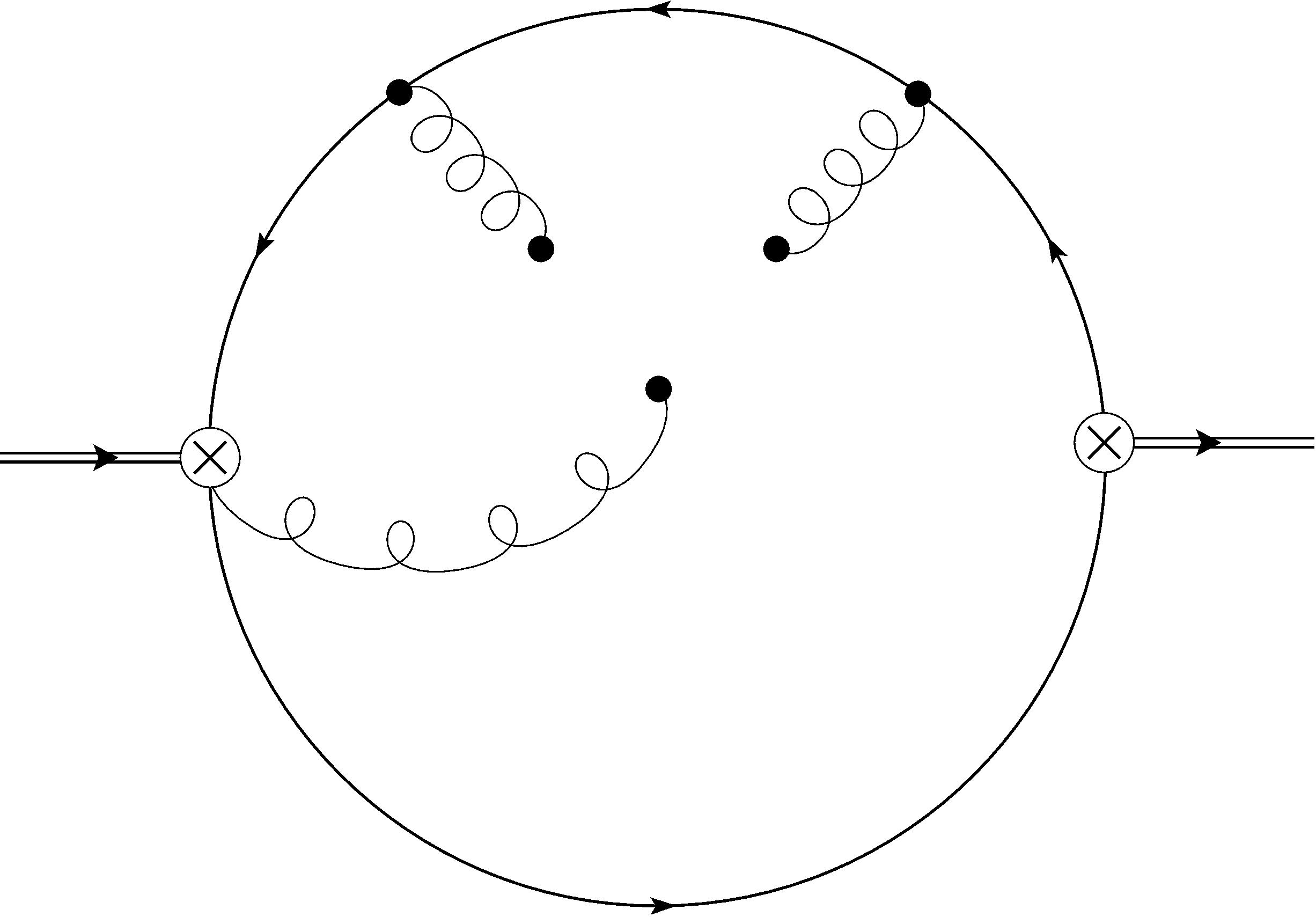} & \includegraphics[width=50mm]{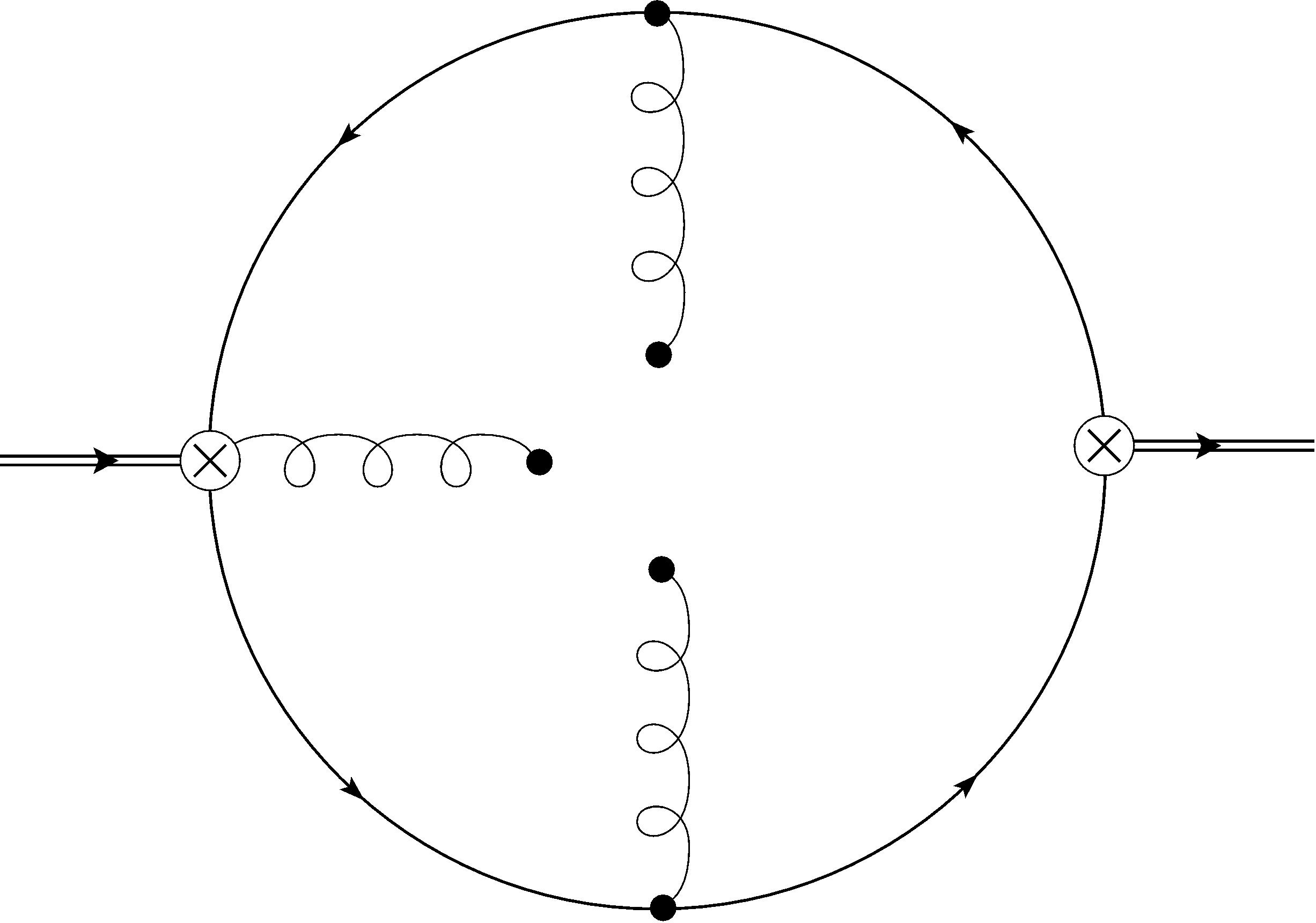} & \includegraphics[width=50mm]{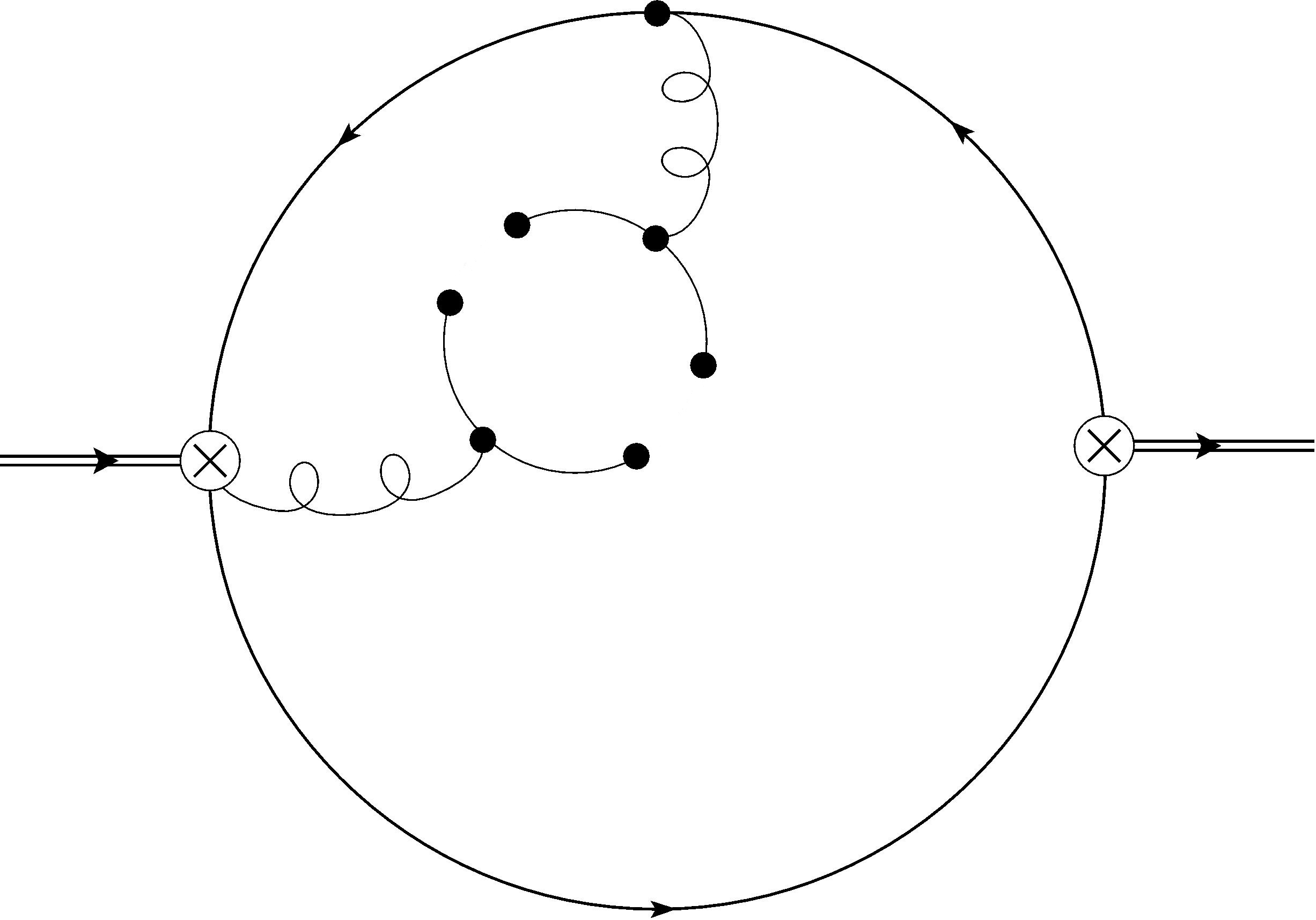}\\
Diagram \rom{4} & Diagram \rom{5} & Diagram \rom{6} \\[15pt]
\end{tabular}
\caption{The LO Feynman diagrams that contribute to the cross-correlator~(\ref{CorFn}) 
which we decompose in~(\ref{fullCorForm}).}
\label{fig01}
\end{figure}

We calculate the correlator~(\ref{CorFn}) within the OPE
in which perturbation theory is supplemented by non-perturbative terms,
each of which is the product of a perturbatively computed Wilson coefficient 
and a nonzero vacuum expectation value, i.e., a QCD condensate.
In addition to perturbation theory, we include OPE terms proportional to the 4d and 6d gluon
condensates and the 6d quark condensate defined respectively by
\begin{equation} \label{fourDcond}
\big\langle \alpha G^{2} \big\rangle = \alpha_{s} \big\langle \! \colon \! G_{\omega\phi}^{a} G_{\omega\phi}^{a} \! \colon\! \big\rangle
\end{equation}
\begin{equation} \label{sixDcond}
\begin{aligned}
\big\langle g^{3}G^{3}\big\rangle = g_{s}^{3} f^{abc} \big\langle \! \colon \! G^{a}_{\omega\zeta} \; G^{b}_{\zeta\rho} G^{c}_{\rho\omega} \! \colon\! \big\rangle
\end{aligned}
\end{equation}
\begin{equation} \label{sixDQcondTr}
\big\langle J^2 \big\rangle =
\mathrm{Tr} \big(\big\langle \! \colon\! J_{\nu} J_{\nu} \! \colon\! \big\rangle \big)
\end{equation}
where
\begin{equation} \label{jDef}
J_{\nu} 
= \frac{-i g_s^2}{4} \lambda^a
\sum_{A} \overline{q}^A \lambda^a \gamma_{\nu} q^A. \end{equation}
In~(\ref{jDef}), the sum on the right-hand side is over quark flavours.
We use the vacuum saturation hypothesis~\cite{Shifman:1978bx} to express
$\langle J^2 \rangle$ in terms of the 3d quark condensate
\begin{equation} \label{threeDcond}
\big\langle \overline{q}q \big\rangle = \big\langle \! \colon \! \overline{q}^{\alpha}_{i} q^{\alpha}_{i} \! \colon\! \big\rangle
\end{equation}
resulting in
\begin{equation} \label{sixDQcond}
\big\langle J^2 \big\rangle
= \frac{2}{3} \, \kappa \, g_s^4 \big\langle \overline{q}q \big\rangle^2
\end{equation}
where $\kappa$ quantifies deviations from exact vacuum saturation. 
Throughout, we set $\kappa = 2$ (see, e.g., ref.~\cite{narisonbook:2004} and references cited therein).
The diagrams that contribute to~(\ref{CorFn}) at LO in the coupling
$g_s$ are shown 
in Figure~\ref{fig01} where we have suppressed a second set of similar diagrams in which the quark 
line runs clockwise.
Wilson coefficients are computed using the fixed-point gauge 
method (see~\cite{PascualTarrach1984,BaganAhmadyEliasEtAl1994}, for example), 
and divergent integrals are handled using dimensional regularization in $D=4+2\epsilon$ dimensions
at $\overline{\text{MS}}$ renormalization scale $\mu$.
As in~\cite{AkyeampongDelbourgo1973}, we use the following convention 
for a dimensionally regularized $\gamma^5$:
\begin{equation} \label{gamma5conv}
\gamma^{5}=\frac{i}{24}\epsilon_{\mu\nu\sigma\rho}\gamma^{\mu}\gamma^{\nu}\gamma^{\sigma}\gamma^{\rho}.
\end{equation}
We employ TARCER~\cite{MertigScharf1998}, 
a Mathematica package that implements the recurrence algorithm of~\cite{Tarasov1996,Tarasov1997},
to express dimensionally regularized integrals  
in terms of a small set of master integrals.
An exact calculation of each needed master integral is either found
in~\cite{BoosDavydychev1991,BroadhurstFleischerTarasov1993} or is a well-known one-loop result.
We denote the OPE computation of $\Pi$ from~(\ref{CorFnProj}) as
$\Pi^{(\text{OPE})}$ which we then decompose as 
\begin{equation} \label{fullCorForm}
  \Pi^{(\text{OPE})} = \Pi^{(\mathrm{\rom{1}})} + \Pi^{(\mathrm{\rom{2}})} + \Pi^{(\mathrm{\rom{3}})} + \Pi^{(\mathrm{\rom{4}})} + \Pi^{(\mathrm{\rom{5}})} + \Pi^{(\mathrm{\rom{6}})}
\end{equation}
where the superscripts in~(\ref{fullCorForm}) correspond to the labels of the diagrams in Figure~\ref{fig01}. 
For $\Pi^{(\text{I})}$, we find an exact $\epsilon$-dependent result
\begin{multline} \label{pertPreXp}
  \Pi^{(\mathrm{\rom{1}})}(z ; \epsilon) = -\frac{\alpha_s \; e^{-2 \epsilon} m^{4(1+\epsilon)} \Gamma(-\epsilon)}{3 \pi^3(3+2 \epsilon)(4 \pi)^{2 \epsilon}}  \Bigg(
  (1 + 2 \epsilon + 4 z (1 + \epsilon)) \Gamma(-\epsilon) \:\hypgeom{3}{2}\left(1,-1-2\epsilon,-\epsilon;\frac{1}{2}-\epsilon,2+\epsilon;z\right)\\
   + \frac{\pi (1+2 \epsilon) \mathrm{Csc}(\pi \epsilon)}{\Gamma(1+\epsilon)} \Bigg(
  -4 + 3 (1 + 4 z (1 + \epsilon)) \:\hypgeom{2}{1}\left(1,-\epsilon;\frac{3}{2};z\right)\\
   -2 (z - 1) \:\hypgeom{3}{2}\left(1,-2\epsilon,-\epsilon;\frac{1}{2}-\epsilon,2+\epsilon;z\right)
  \Bigg)
  \Bigg)
\end{multline}
where
\begin{equation}
  z=\frac{q^2}{4m^2},
\end{equation}
$m$ is a heavy quark mass (i.e., $m_c$ or $m_b$), $\Gamma$ is the gamma function, and 
$\hypgeom{p}{q}(\cdots;\cdots;z)$ are generalized hypergeometric functions 
(see~\cite{AbramowitzStegun1965}, for example). 
Expanding~(\ref{pertPreXp}) in $\epsilon$ and dropping terms polynomial in $z$ as they will not contribute to the LSR, we find
\begin{equation} \label{pertXpNow}
\Pi^{(\mathrm{\rom{1}})}(z) =\frac{2\alpha_s m^4 z(1+4z)\:\hypgeom{2}{1}\left(1,1;\frac{5}{2};z\right)}{9\pi^3}\frac{1}{\epsilon}
+\frac{d}{d\epsilon}\Pi^{(\mathrm{\rom{1}})}(z ; \epsilon) \Big|_{\epsilon = 0}.
\end{equation}
For the sake of brevity, we do not include an explicit expression for the derivative
term on the right-hand side of~(\ref{pertXpNow}).
(Note that~(\ref{pertXpNow}) is ultimately superseded by~(\ref{expRenormedPert}),
and we provide a complete expression for the latter.)
Expanding the remaining terms on the right hand side of~(\ref{fullCorForm}) in $\epsilon$, we find
\begin{gather}
\Pi^{(\mathrm{\rom{2}})}(z) = \frac{z\Big(-3+\hypgeom{2}{1}\big(1,1;\frac{5}{2};z\big)\Big)}{18\pi(z-1)}
\big\langle\alpha G^{2} \big\rangle
\label{expfourd}\\
\Pi^{(\mathrm{\rom{3}})}(z) = \frac{\Big(2+5z-4z^2-(2-7z+10z^2-4z^3)\:\hypgeom{2}{1}\big(1,1;\frac{5}{2};z\big)\Big)}{2304\pi^2 m^2 (z-1)^3}
\big\langle g^{3} G^{3} \big\rangle
\label{expsixd}\\
\Pi^{(\text{\rom{4}})}(z) = \frac{\big\langle g^{3} G^{3} \big\rangle}{4608\pi^2 m^2 (z-1)^3}\Bigg( -22+41z-16z^2+(10-25z+22z^2-8z^3)\:
\hypgeom{2}{1}\big(1,1;\textstyle{\frac{5}{2}};z\big)\Bigg)
\label{expDiag4}\\
\Pi^{(\text{\rom{5}})}(z) = \frac{\big\langle g^{3} G^{3} \big\rangle}{4608\pi^2 m^2 (z-1)^2}\Bigg(-15+12z+(3-2z)\:
\hypgeom{2}{1}\big(1,1;\textstyle{\frac{5}{2}};z\big)\Bigg)
\label{expDiag5}\\
\Pi^{(\text{\rom{6}})}(z) = \frac{2 \alpha_s^2 \big\langle \overline{q}q \big\rangle^2}{81 m^2 (z-1)^3}\Bigg( 2 +5z -4z^2 +(-2+7z-10z^2+4z^3)\:
\hypgeom{2}{1}\big(1,1;\textstyle{\frac{5}{2}};z\big)\Bigg).
\label{expDiag6}
\end{gather}

\begin{figure}
\centering
\begin{tabular}{cc}
\includegraphics[width=50mm]{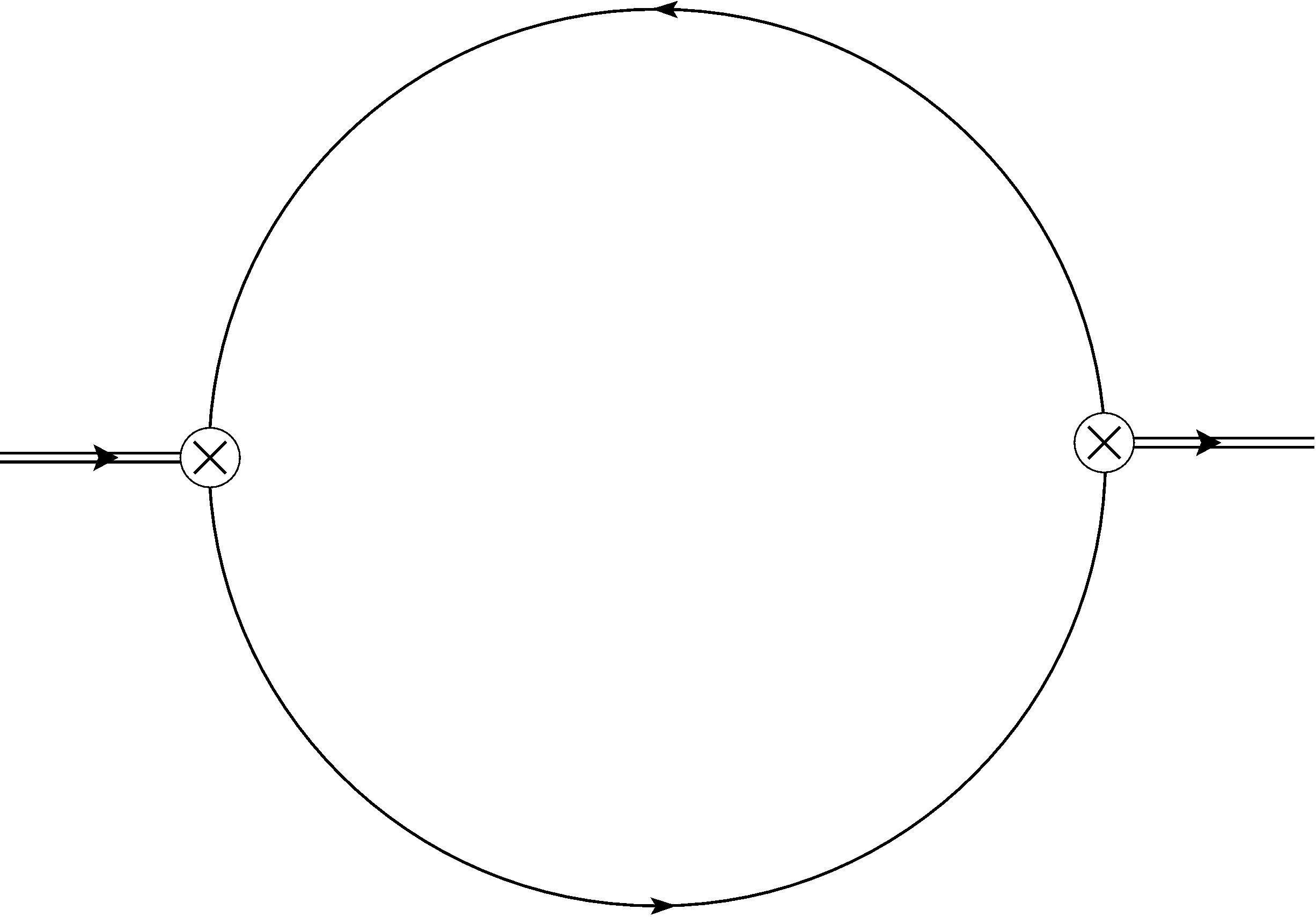} & \includegraphics[width=50mm]{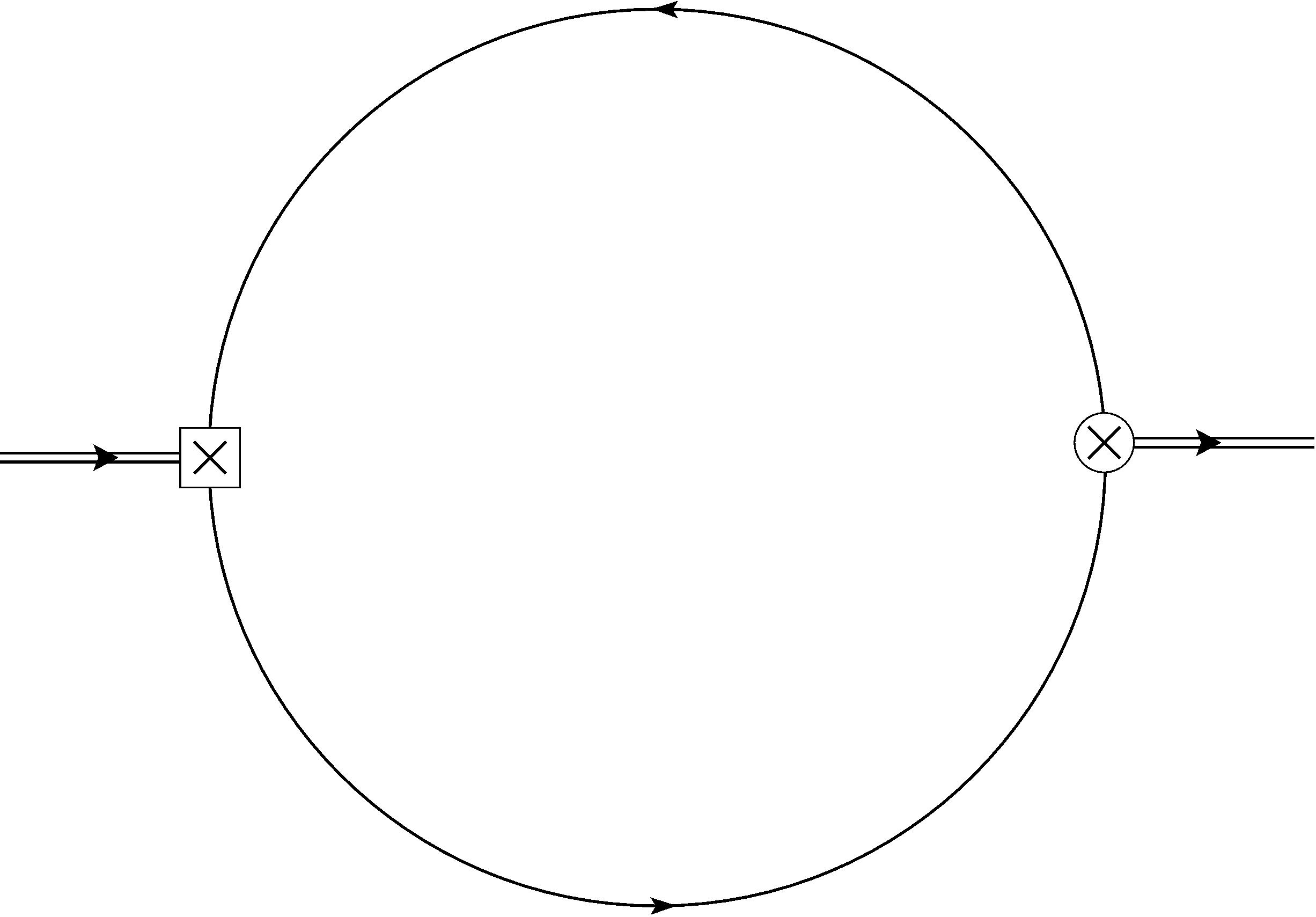} \\
Diagram R\rom{1} & Diagram R\rom{2} \\[15pt]
\end{tabular}
\caption{Renormalization-induced Feynman diagrams that provide a LO perturbative contribution
to the mixed correlator. The square insertion denotes the current~(\ref{DCurrent}).}
\label{fig02}
\end{figure}

Perturbation theory~(\ref{pertXpNow}) contains a nonlocal divergence.
Following~\cite{Chen:2013pya,ho:2017}, this divergence is eliminated
via operator mixing under renormalization. The meson current~(\ref{CurMes}) is renormalization-group (RG) invariant and so we only need consider the operator mixing of the hybrid current~(\ref{CurHyb})
which induces operator mixing  with~(\ref{CurMes}) and with
\begin{equation} \label{DCurrent}
j_{\nu}^{(c)} = \overline{Q} i D_{\nu} Q
\end{equation}
where $D_{\nu}=\partial_{\nu}-\frac{i}{2} g_s \lambda^a A^a_{\nu}$ is the covariant
derivative.  Thus, 
\begin{equation} \label{renormTheLast}
  j_{\nu}^{(\text{h})}  \rightarrow  j_{\nu}^{(\text{h})} 
  + \frac{C_{1}}{\epsilon}j_{\nu}^{(\text{m})} 
  + \frac{C_{2}}{\epsilon}j_{\nu}^{(\text{c})}
\end{equation}
where $C_1$ and $C_2$ are as-yet-undetermined 
renormalization constants.
Substituting~(\ref{renormTheLast}) into~(\ref{CorFn}) 
(in $D$ rather than four dimensions) gives
\begin{multline} \label{renormBig}
   i\! \int \!\! d^{D}\!x \; e^{iq\cdot x} \langle\Omega | 
   \tau j_{\mu}^{(\text{m})} j_{\nu}^{(\text{h})} |\Omega\rangle
   \rightarrow 
   i\! \int \!\! d^{D}\!x \; e^{iq\cdot x} \langle\Omega|\tau j_{\mu}^{(\text{m})} 
   j_{\nu}^{(\text{h})} |\Omega\rangle \\
   +i\frac{C_{1}}{\epsilon} \int \!\! d^{D}\!x \; e^{iq\cdot x}\langle\Omega|\tau j_{\mu}^{(\text{m})}
   j_{\nu}^{(\text{m})} |\Omega\rangle 
   +i\frac{C_{2}}{\epsilon}\int\!\! d^{D}\!x \; e^{iq\cdot x} \langle\Omega|\tau j_{\mu}^{(m)}
   j_{\nu}^{(\text{c})} | \Omega \rangle.
\end{multline}
The last two terms on the right-hand side of~(\ref{renormBig}) 
each generate a new renormalization-induced 
Feynman diagram, the pair of which are shown in Figure~\ref{fig02}. 
Note that a square insertion represents the current~(\ref{DCurrent}).
Evaluating these two diagrams and choosing $C_1$ and $C_2$ 
such that the right-hand side of~(\ref{renormBig}) is free of nonlocal divergences, we find
\begin{gather}
C_{1} = -\frac{10 m^2 \alpha_s}{9 \pi}\\ 
C_{2} = \frac{4 m \alpha_s}{9 \pi}
\end{gather}
as well as an updated expression for $\Pi^{(\text{\rom{1}})}$ from~(\ref{fullCorForm}) that is free of nonlocal divergences
\begin{multline} \label{expRenormedPert}
  \Pi^{(\text{\rom{1}})}(z) = \frac{2 \alpha_s m^4 z}{81 \pi^3} 
  \Bigg(18(z-1)\:\hypgeom{3}{2}\big(1,1,1;\tfrac{3}{2},3;z\big) -2z (4 z + 1)\:
  \hypgeom{3}{2}\big(1,1,2;\tfrac{5}{2},4;z\big)\\
  +3 \bigg(3 (4 z+1) \log \left(\frac{m^2}{\mu ^2}\right)+26 z+6 \bigg)\:
  \hypgeom{2}{1}\big(1,1;\tfrac{5}{2};z\big) \Bigg)
\end{multline}
where, again, we have omitted polynomials in $z$ as they will not contribute to the LSR.

In summary, taking operator mixing into account, the LO QCD expression 
$\Pi^{(\text{OPE})}$ can be decomposed as in~(\ref{fullCorForm}) with
the terms on the right-hand side given by~(\ref{expRenormedPert}) 
and~(\ref{expfourd})--(\ref{expDiag6}).

%%%%%%%%%% QCD Laplace Sum-Rules %%%%%%%%%%
\section{QCD Laplace Sum-Rules}\label{III}
The function $\Pi$ from~(\ref{CorFnProj}) satisfies a dispersion relation
\begin{equation}\label{dispersion_relation}
  \Pi(Q^2)=\frac{Q^6}{\pi}\int_{t_0}^{\infty}
  \frac{\mathrm{Im}\Pi(t)}{t^3(t+Q^2)}
  \dif{t} +\cdots,\ Q^2=-q^2>0
\end{equation}
where $\Pi$ on the left-hand side is to be identified with the QCD prediction $\Pi^{\text{(OPE)}}$;
$\mathrm{Im}\Pi(t)$ is the hadronic spectral function;
$t_0$ is the hadron threshold parameter;
and $\cdots$ represents subtraction constants, collectively a quadratic polynomial in $Q^2$.
To eliminate these subtraction constants as well as local divergences in 
$\Pi^{\text{(OPE)}}$ 
and to accentuate the resonance contributions of the hadronic spectral function
to the integral on the right-hand side of~(\ref{dispersion_relation}),
we apply the Borel transform 
\begin{equation}\label{borel}
  \hat{\mathcal{B}}=\!\lim_{\stackrel{N,Q^2\rightarrow\infty}{\tau=N/Q^2}}
  \!\frac{\big(-Q^2\big)^N}{\Gamma(N)}\bigg(\frac{d}{dQ^2}\bigg)^N
\end{equation}
with Borel parameter $\tau$
to formulate the $0^{\text{th}}$-order LSR~\cite{Shifman:1978bx}
\begin{equation} \label{zeroThLSR}
\lsr(\tau)\equiv\frac{1}{\tau}\hat{\mathcal{B}}\Big\{\Pi(Q^2) \Big\}\\
= \int_{t_0}^{\infty} e^{-t\tau}\frac{1}{\pi}\mathrm{Im}\Pi(t)\dif{t}.
\end{equation}
On the right-hand side of~(\ref{zeroThLSR}), we use a 
``resonance(s) plus continuum'' model
\begin{equation} \label{ResCont}
\frac{1}{\pi}\mathrm{Im}\Pi(t)
    =\rho^{\text{(had)}}(t)+\frac{1}{\pi}\mathrm{Im}\Pi^{\text{(OPE)}}(t)\theta(t-s_0)
\end{equation}
where 
$\rho^{\text{(had)}}$ represents the resonance content of the spectral function
(to be discussed further in Section~\ref{IV}),
$\theta$ is the Heaviside step function, and $s_0$ is the continuum threshold.
Then, we define the continuum-subtracted $0^{\text{th}}$-order LSR
\begin{equation} \label{subedLSR}
\lsr(\tau, s_0) \equiv \lsr(\tau) - \int_{s_0}^{\infty} e^{-t\tau}\frac{1}{\pi}\mathrm{Im}\Pi^{\text{(OPE)}}(t)\dif{t}\\
=\int_{t_0}^{s_0} e^{-t\tau} \rho^{\text{(had)}}(t)\dif{t}.
\end{equation}

\begin{figure}
\centering
\includegraphics[scale=0.8]{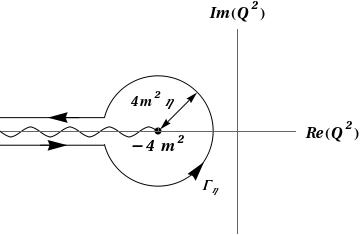}
\caption{\label{keyhole} The integration contour used to compute the 
  LSR~(\ref{intermediateTheoryLSR})}
\end{figure}

To compute $\lsr\double{\tau}{s_0}$,
we use the following identity relating the Borel transform 
to the inverse Laplace transform $\hat{\mathcal{L}}^{-1}$~\cite{Shifman:1978bx}:
\begin{equation}\label{borelIdentity}
\begin{aligned}
  \frac{1}{\tau}\hat{\mathcal{B}} \Big\{ f(Q^2) \Big\}&=\hat{\mathcal{L}}^{-1} \Big\{ f(Q^2) \Big\}\\
  &= \frac{1}{2 \pi i} \int_{c-i \infty}^{c+i \infty} f(Q^2) e^{Q^2 \tau} \dif{Q^2}
\end{aligned}
\end{equation}
where $c$ is any real number for which $f(Q^2)$ is analytic for 
$\text{Re}(Q^2)>c$. 
Generalized hypergeometric functions of the form $\hypgeom{p}{p-1}$ have a branch cut along
the positive real semi-axis originating at the branch point $z=1$.
As such, in the complex $Q^2$-plane, $\Pi^{\text{(OPE)}}(Q^2)$ is analytic except 
for a branch cut along the negative real semi-axis originating at a branch 
point $Q^2=-4m^2$.
In~(\ref{borelIdentity}), we let $f(Q^2)=\Pi^{\text{(OPE)}}(Q^2)$
and deform the integration contour on the right-hand side to that shown in
Figure~\ref{keyhole}.  
Then, we apply definitions~(\ref{zeroThLSR}) and~(\ref{subedLSR}) 
to find
\begin{equation}\label{intermediateTheoryLSR}
  \lsr\double{\tau}{s_0}=\int_{4m^2(1+\eta)}^{s_0}e^{-t\tau}\frac{1}{\pi}
    \text{Im}\Pi^{\text{(OPE)}}(t)\dif{t}
  +\frac{1}{2\pi i}\int_{\Gamma_{\eta}} e^{Q^2 \tau}
    \Pi^{\text{(OPE)}}(Q^2) \dif{Q^2}\ \;\; \text{for}\ \;\; \eta\rightarrow0^{+}
\end{equation}
where
\begin{equation}\label{breakdown}
  \text{Im}\Pi^{\text{(OPE)}}(t)=
  \sum_{i=\rom{1}}^{\rom{6}}\text{Im}\Pi^{\text{(i)}}(t)
\end{equation}
and, from~(\ref{expRenormedPert}) and~(\ref{expfourd})--(\ref{expDiag6}), we get
\begin{gather} \label{ImP}
\begin{split}
\mathrm{Im}\Pi^{\text{(\rom{1})}}(t) = 
\frac{\alpha_s}{18\pi^2 t\sqrt{t-4m^2}}
\Bigg(
  24m^3\sqrt{\frac{t}{4m^2}-1}\,\big(2m^4-2m^2 t+t^2\big)
  \sinh^{-1}\bigg(\sqrt{\frac{t}{4m^2}-1}\bigg)
  \\
  +\sqrt{t}\,\big(t-4m^2\big)\Big(18m^4+8m^2 t-t^2+6m^2(t+m^2)\Big)\log\bigg(\frac{m^2}{\mu^2}\bigg)
\Bigg)
\end{split}
\\
\label{Imfourd}
  \mathrm{Im}\Pi^{\text{(\rom{2})}}(t)
  = \frac{m^2}{3\sqrt{t(t-4m^2)}}\glueFourD
\\
\label{Imsixd}
  \mathrm{Im}\Pi^{\text{(\rom{3})}}(t)=
  \frac{t^3 - 10m^2 t^2 + 28m^4 t - 32m^6}{96 \pi t^{3/2} (t-4m^2)^{5/2}}
  \glueSixD
  \\
\label{ImDiag4}
  \mathrm{Im}\Pi^{\text{(\rom{4})}}(t)=
  \frac{-t^3 + 11m^2 t^2 - 50m^4 t + 80m^6}{96 \pi t^{3/2} (t-4m^2)^{5/2}}
  \glueSixD
  \\
\label{ImDiag5}
  \mathrm{Im}\Pi^{\text{(\rom{5})}}(t)=
  \frac{-t + 6m^2}{96 \pi t^{3/2} (t-4m^2)^{3/2}}
  \glueSixD
  \\
\label{ImDiag6}
  \mathrm{Im}\Pi^{\text{(\rom{6})}}(t)=
  \frac{16 \pi \alpha_s^2 (t^3 - 10m^2 t^2 + 28m^4 t - 32m^6)}{27 t^{3/2} (t-4m^2)^{5/2}}
  \big\langle \overline{q}q \big\rangle^2.
\end{gather}

For both $\Pi^{\text{(\rom{1})}}$ and $\Pi^{\text{(\rom{2})}}$, 
the first integral on the right-hand side 
of~(\ref{intermediateTheoryLSR}) converges and the second vanishes
for $\eta\rightarrow 0^{+}$.
For $\Pi^{\text{(\rom{3})}}$--$\Pi^{\text{(\rom{6})}}$,
however, each integral diverges although their sum is finite.
To isolate this finite contribution, 
we first expand the imaginary parts~(\ref{Imsixd})--(\ref{ImDiag6}) near $t=4m^2$:
\begin{multline}\label{expansion}
  \text{Im}\Pi^{\text{(\rom{3})}}(t) + \text{Im}\Pi^{\text{(\rom{4})}}(t) + \text{Im}\Pi^{\text{(\rom{5})}}(t) + \text{Im}\Pi^{\text{(\rom{6})}}(t)=\\
  \frac{-m}{864\pi\sqrt{t-4m^2}} \Bigg(
   \frac{m^2 \Big(27 \glueSixD + 1024 \pi^2 \alpha_s^2 \big\langle \overline{q}q \big\rangle^2 \Big)}{(t-4m^2)^2} + 
  \frac{27 \glueSixD - 1024 \pi^2 \alpha_s^2 \big\langle \overline{q}q \big\rangle^2}{8(t-4m^2)}
  + p(t)\Bigg)
\end{multline}
where 
\begin{multline}\label{pExplicit}
p(t)= \frac{-27}{8 (2m + \sqrt{t})^2 t^{3/2}} 
\bigg( 
16 m^3 + 16 m^2 \sqrt{t} + 4 m t + t^{3/2}
\bigg) 
\glueSixD \\
+ \frac{1024 \pi^2 \alpha_s^2}{8m (2m + \sqrt{t})^2 t^{3/2}} 
\bigg( 
32 m^4 + 32 m^3 \sqrt{t} - 4 m^2 t - 15 m t^{3/2} - 4 t^2
\bigg)  
\big\langle \overline{q}q \big\rangle^2
\end{multline}
is analytic in a neighbourhood about $t=4m^2$.
When~(\ref{expansion}) is inserted into the first integral on the right-hand side 
of~(\ref{intermediateTheoryLSR}),
the part of the result stemming from the $p(t)$ term
converges whereas the parts stemming from the $(t-4m^2)^{-2}$ 
and $(t-4m^2)^{-1}$ terms diverge.
Focusing on these divergent parts, we have
\begin{multline}\label{firstIntegral}
  \int_{4m^2(1+\eta)}^{s_0}e^{-t\tau}\frac{1}{\pi}
    \Bigg(\frac{-m}{864\pi\sqrt{t-4m^2}} \bigg(
  \frac{m^2 \Big(27 \glueSixD + 1024 \pi^2 \alpha_s^2 \big\langle \overline{q}q \big\rangle^2 \Big)}{(t-4m^2)^2}\\
 + \frac{27 \glueSixD - 1024 \pi^2 \alpha_s^2 \big\langle \overline{q}q \big\rangle^2}{8(t-4m^2)}
  \bigg)\Bigg)
    \dif{t}
    \\
=\frac{e^{-4m^2 \tau}}{10368\pi^2}
  \Bigg(-\frac{27 \glueSixD + 1024 \pi^2 \alpha_s^2 \big\langle \overline{q}q \big\rangle^2}{\eta^{3/2}}\\
  +\frac{3\Big(27 (8m^2 \tau -1) \glueSixD + 1024 (8m^2 \tau +1) \pi^2 \alpha_s^2 \big\langle \overline{q}q \big\rangle^2 \Big)}{2 \eta^{1/2}} \Bigg)\\
 +\frac{m \; e^{-4m^2 \tau}}{384\pi^2}
  \Bigg(
  \sqrt{\pi\tau} \Big( 3 - 8m^2 \tau \Big) \mathrm{erf}\Big(\sqrt{(s_0-4m^2)\tau}\Big) + \frac{e^{-s_0 \, \tau}}{(s_0 - 4m^2)^{3/2}} \bigg( \\
  -8 e^{s_0 \, \tau} m^2 \sqrt{\pi} \big((s_0 - 4m^2) \tau \big)^{3/2} + e^{4m^2 \, \tau} \big( 3 s_0 +32 m^4 \tau -8 m^2 (1+s_0 \tau)\big) \\
  +6 e^{s_0 \, \tau} m^2 \, \mathrm{E}_{5/2}\big( (s_0 -4m^2) \tau \big)
  \bigg)
  \Bigg)\glueSixD \\
 +\frac{8 \; \alpha_s \; m \; e^{-4m^2 \tau}}{81}
  \Bigg(
- \sqrt{\pi\tau} \Big( 3 - 8m^2 \tau \Big) \mathrm{erf}\Big(\sqrt{(s_0-4m^2)\tau}\Big) + \frac{e^{-s_0 \, \tau}}{(s_0 - 4m^2)^{3/2}} \bigg( \\
  -24 e^{s_0 \, \tau} m^2 \sqrt{\pi} \big((s_0 - 4m^2) \tau \big)^{3/2} + e^{4m^2 \, \tau} \big( -3 s_0 +8 m^2 (1 - 4m^2 \tau +s_0 \tau)\big) \\
   +18 e^{s_0 \, \tau} m^2 \, \mathrm{E}_{5/2}\big( (s_0 -4m^2) \tau \big) 
  \bigg)
  \Bigg)\big\langle \overline{q}q \big\rangle^2
\end{multline}
for $\eta\rightarrow 0^{+}$. In~(\ref{firstIntegral}), 
$\mathrm{erf}$ is the error function and $\mathrm{E}_{n}$ is the exponential integral function
\begin{equation}
  \mathrm{erf}(z)=\frac{2}{\sqrt{\pi}}\int_0^z e^{-t^2}\dif{t}
\end{equation}
\begin{equation}
  \mathrm{E}_{n}(z)=\int_1^\infty \frac{e^{-z t}}{t^n}\dif{t}.
\end{equation}
On the right-hand side of~(\ref{firstIntegral}), note that the terms proportional to $\eta^{-3/2}$ 
and $\eta^{-1/2}$ diverge whereas the remaining terms are finite.
Next, we consider the contributions of $\Pi^{\text{(\rom{3})}}$--$\Pi^{\text{(\rom{6})}}$
to the second integral on the right-hand side of~(\ref{intermediateTheoryLSR}).
Parameterizing 
\begin{equation}
  Q^2=-4m^2+4m^2 \eta e^{i\theta}
\end{equation}
for $\theta_i=-\pi^{+}$ to $\theta_f=\pi^{-}$, we find that
\begin{multline}\label{secondIntegral}
  \frac{1}{2\pi i}\int_{\Gamma_{\eta}} e^{Q^2 \tau} \Big(
    \Pi^{\text{(\rom{3})}}(Q^2) + \Pi^{\text{(\rom{4})}}(Q^2) + \Pi^{\text{(\rom{5})}}(Q^2) + \Pi^{\text{(\rom{6})}}(Q^2) \Big) \dif{Q^2} \\
  = - \frac{e^{-4m^2 \tau}}{10368\pi^2}
  \Bigg(-\frac{27 \glueSixD + 1024 \pi^2 \alpha_s^2 \big\langle \overline{q}q \big\rangle^2}{\eta^{3/2}} \\
   +\frac{3\Big(27 (8m^2 \tau -1) \glueSixD + 1024 (8m^2 \tau +1) \pi^2 \alpha_s^2 \big\langle \overline{q}q \big\rangle^2 \Big)}{2 \eta^{1/2}}\Bigg)
  + \frac{e^{-4m^2 \tau}}{384\pi^2} \glueSixD
\end{multline}
for $\eta\rightarrow 0^{+}$.
When~(\ref{firstIntegral}) and~(\ref{secondIntegral}) are added together,
the divergent terms which go like $\eta^{-3/2}$ and $\eta^{-1/2}$ cancel
leaving a finite result.
Finally, collecting together~(\ref{intermediateTheoryLSR}),
(\ref{breakdown}), (\ref{expansion}), (\ref{firstIntegral}),
and~(\ref{secondIntegral}), we have
\begin{multline}\label{finalTheoryLSR}
  \lsr\double{\tau}{s_0}=\int_{4m^2}^{s_0}e^{-t\tau}\frac{1}{\pi}
    \Bigg(\text{Im}\Pi^{\text{(\rom{1})}}(t)+\text{Im}\Pi^{\text{(\rom{2})}}(t)
-\frac{m \; p(t)}{864\pi\sqrt{t-4m^2}} 
  \Bigg)
    \dif{t}
    \\
 +\frac{m \; e^{-4m^2 \tau}}{384\pi^2}
  \Bigg(
  \frac{1}{m} + \sqrt{\pi\tau} \Big( 3 - 8m^2 \tau \Big) \mathrm{erf}\Big(\sqrt{(s_0-4m^2)\tau}\Big) + \frac{e^{-s_0 \, \tau}}{(s_0 - 4m^2)^{3/2}} \bigg( \\
  -8 e^{s_0 \, \tau} m^2 \sqrt{\pi} \big((s_0 - 4m^2) \tau \big)^{3/2} + e^{4m^2 \, \tau} \big( 3 s_0 +32 m^4 \tau -8 m^2 (1+s_0 \tau)\big) \\
 +6 e^{s_0 \, \tau} m^2 \, \mathrm{E}_{5/2}\big( (s_0 -4m^2) \tau \big)
  \bigg)
  \Bigg)\glueSixD \\
 +\frac{8 \; \alpha_s \; m \; e^{-4m^2 \tau}}{81}
  \Bigg(
- \sqrt{\pi\tau} \Big( 3 - 8m^2 \tau \Big) \mathrm{erf}\Big(\sqrt{(s_0-4m^2)\tau}\Big) + \frac{e^{-s_0 \, \tau}}{(s_0 - 4m^2)^{3/2}} \bigg( \\
  -24 e^{s_0 \, \tau} m^2 \sqrt{\pi} \big((s_0 - 4m^2) \tau \big)^{3/2} + e^{4m^2 \, \tau} \big( -3 s_0 +8 m^2 (1 - 4m^2 \tau +s_0 \tau)\big) \\
   +18 e^{s_0 \, \tau} m^2 \, \mathrm{E}_{5/2}\big( (s_0 -4m^2) \tau \big) 
  \bigg)
  \Bigg)\big\langle \overline{q}q \big\rangle^2
\end{multline}
where, again, $p(t)$ is given in~(\ref{pExplicit}), 
and the imaginary parts $\mathrm{Im}\Pi^{\text{(I)}}$ and 
$\mathrm{Im}\Pi^{\text{(II)}}$ are given
in~(\ref{ImP}) and~(\ref{Imfourd}).  
The integral on the right-hand side 
of~(\ref{finalTheoryLSR}) can be evaluated analytically; however, the result
is long and so we omit it for the sake of brevity.

Renormalization-group improvement~\cite{Narison:1981ts}
implies that the strong coupling and 
quark mass in the simplified~(\ref{intermediateTheoryLSR}) get replaced by 
corresponding running quantities evaluated at renormalization scale $\mu$, 
i.e., $\alpha_s\rightarrow\alpha_s(\mu)$ 
and $m\rightarrow m_{c,b}(\mu)$.
At one-loop in the $\overline{\text{MS}}$ renormalization scheme, we have
for charmonium
\begin{gather}
  \alpha_s(\mu) = \frac{\alpha_s(M_{\tau})}{1 + \frac{25 \alpha_s(M_{\tau})}%
  {12\pi}\log\!{\Big(\frac{\mu^2}{M_{\tau}^2}\Big)}}
  \\ 
  m_{c}(\mu) = \overline{m}_{c}\bigg(\frac{\alpha_s(\mu)}
  {\alpha_s(\overline{m}_{c})}\bigg)^{12/25}
\end{gather}
and for bottomonium
\begin{gather}
  \alpha_s(\mu) = \frac{\alpha_s(M_Z)}{1 + \frac{23 \alpha_s(M_Z)}%
  {12\pi}\log\!{\Big(\frac{\mu^2}{M_Z^2}\Big)}}
  \\ 
  m_{b}(\mu) = \overline{m}_{b}\bigg(\frac{\alpha_s(\mu)}
  {\alpha_s(\overline{m}_{b})}\bigg)^{12/23}
\end{gather}
where~\cite{Olive:2016xmw}
\begin{gather}
  \alpha_s(M_{\tau})=0.330\pm0.014\label{alphatau}\\
  \alpha_s(M_{Z})=0.1185\pm0.0006\label{alphaZ}\\
  \overline{m}_c=(1.275\pm0.025)\ \text{GeV}\label{charmMass}\\
  \overline{m}_b=(4.18\pm 0.03)\ \text{GeV}\label{bottomMass}.
\end{gather}
For charmonium, we set $\mu$ to $\overline{m}_c$;
for bottomonium, we set $\mu$ to $\overline{m}_b$.
Finally, we use the following values for the gluon and quark
condensates~\cite{Launer:1983ib,Narison2010,ChenKleivSteeleEtAl2013}:
\begin{gather}
  \glueFourD=(0.075\pm0.02)\ \text{GeV}^4\label{glueFourDValue}\\
  \glueSixD=((8.2\pm1.0)\ \text{GeV}^2)\label{glueSixDValue}\glueFourD\\
 \big\langle \overline{q}q \big\rangle = -(0.23 \pm 0.03)^3\ \text{GeV}^3\label{quarkThreeDValue}.
\end{gather}

%%%%%%%%%% Analysis and Results %%%%%%%%%%
\section{Analysis and Results}\label{IV}
To extract hadron properties from the LSR~(\ref{finalTheoryLSR})
we must first select an acceptable range of $\tau$ values, i.e., a Borel interval
$\double{\tau_{\text{min}}}{\tau_{\text{max}}}$.
To do so, we follow the same methodology 
as in~\cite{Chen:2013pya,ho:2017,BergHarnettKleivEtAl2012,Harnett:2012gs}.
To choose $\tau_{\text{max}}$, we demand that the LSR converge
in the following sense: the magnitude of the 
4d gluon condensate contribution (stemming from $\Pi^{\text{(\rom{2})}}$)
must be less than one-third that of the perturbative contribution
(stemming from $\Pi^{\text{(\rom{1})}}$), 
and the magnitude of the sum of the 
6d gluon and quark condensate contributions 
(stemming from $\Pi^{\text{(\rom{3})}}$--$\Pi^{\text{(\rom{6})}}$)
must be less than one-third
that of 4d gluon condensate contribution.
For charmonium, we find $\tau_{\text{max}}=0.6\ \text{GeV}^{-2}$;
for bottomonium, we find $\tau_{\text{max}}=0.2\ \text{GeV}^{-2}$.
To choose $\tau_{\text{min}}$, we consider the pole contribution
\begin{equation}\label{pole_contribution}
  \frac{\lsr\double{\tau}{s_0}}{\lsr\double{\tau}{\infty}},
\end{equation}
i.e., the ratio of the LSR's hadron contribution to its hadron plus continuum contribution, and demand that it be at least 10\%.
In both the charmonium and bottomonium analyses, 
the value of $\tau_{\text{min}}$ selected using this prescription
depends weakly on $s_0$, a parameter not known at the outset.
Hence, we first choose reasonable seed values for $s_0$:
$s_0=25$~GeV$^2$ for charmonium and $s_0=130$~GeV$^2$ for bottomonium.
When input into~(\ref{pole_contribution}), these two seed values correspond to 
$\tau_{\text{min}}=0.1\ \text{GeV}^{-2}$ for charmonium and
$\tau_{\text{min}}=0.01\ \text{GeV}^{-2}$ for bottomonium.
After making predictions for $s_0$ through the optimization procedure explained below, 
we then update $\tau_{\text{min}}$ using the new, predicted value of $s_0$.
In all cases considered, the effect on $\tau_{\text{min}}$ was insignificant.

Next, we turn our attention to $\rho^{\text{(had)}}$ from~(\ref{ResCont}).
As $\rho^{\text{(had)}}$ represents the resonance(s) portion of the hadronic
spectral function, it contains those hadrons which couple to \emph{both} the 
meson current~(\ref{CurMes}) and the hybrid current~(\ref{CurHyb}).
Such hadrons can be thought of as mixtures that have a 
$\overline{q}q$-meson and a hybrid component.
Our analysis approach is to input known vector heavy quarkonium
resonances into $\rho^{\text{(had)}}$ in order to test them for 
meson-hybrid mixing.
In Table~\ref{charmoniumTable}, we list all vector charmonium
resonances that have a Particle Data Group entry in~\cite{Olive:2016xmw}, and 
in Table~\ref{bottomoniumTable}, we do the same for bottomonium.
(Note that, in Table~\ref{charmoniumTable}, states named with a $\psi$ or $J/\psi$
have $I^G=0^{-}$ whereas those named with an $X$ have unknown $I^G$.)
All resonances listed in the two tables have widths $\lesssim100$~MeV.
In general, LSRs are insensitive to resonance widths of up to several 
hundred MeV, and so, we ignore the widths of individual resonances.
But, for a cluster of resonances for which the mass difference between 
successively heavier states is $\lesssim250$~MeV, we amalgamate the cluster
into a single resonance with nonzero effective width.
And so, we consider a variety of $\rho^{\text{(had)}}$ of the form
\begin{equation}\label{rhoForms}
  \rho^{\text{(had)}}(t)=\sum_{i=1}^n \rho^{\text{(had)}}_i(t)
\end{equation}
where $n$ is the number of distinct resonances (or clusters of resonances) and 
where each $\rho_i^{\text{(had)}}$ is either a narrow $(\Gamma_i=0)$ resonance
\begin{equation}\label{narrow}
  \rho_i^{\text{(had)}}(t) = \xi_i \delta(t-m_i^2)
\end{equation}
or, for a resonance cluster, a rectangular pulse 
\begin{equation}\label{pulse}
  \rho_i^{\text{(had)}}(t) = \frac{\xi_i}{2m_i \Gamma_i}
  \theta\big(t-m_i(m_i-\Gamma_i)\big)\theta\big(m_i(m_i+\Gamma_i)-t\big)
\end{equation}
with effective width $\Gamma_i \neq 0$ 
in which the resonance strength is uniformly distributed over 
$m_i(m_i-\Gamma_i)<t<m_i(m_i+\Gamma_i)$.
The $\{\xi_i\}_{i=1}^n$ are mixing parameters related to
the combined effect of coupling to the hybrid and $q\overline{q}$-meson currents. 
A state with both $q\overline{q}$-meson and hybrid components has 
$\xi_i\neq0$; a pure $q\overline{q}$-meson or pure hybrid state has 
$\xi_i=0$.
The specific models for which we present results are defined for the charmonium 
and bottomonium sectors in Tables~\ref{charmoniumModels} and~\ref{bottomoniumModels} respectively.

\begin{table}
\caption{Particle Data Group masses of vector charmonium resonances~\cite{Olive:2016xmw}.}
\label{charmoniumTable}
\centering
\begin{tabular}{lS}
  \addlinespace
  \toprule
  Name &  \multicolumn{1}{c}{Mass (\si{GeV})} \\
  \midrule
  $J/\psi$ & 3.10 \\
  \addlinespace
  $\psi(2S)$ & 3.69 \\
  $\psi(3770)$ & 3.77 \\
  \addlinespace
  $\psi(4040)$ & 4.04 \\
  $\psi(4160)$ & 4.19 \\
  $X(4230)$ & 4.23 \\
  $X(4260)$ & 4.23 \\
  $X(4360)$ & 4.34 \\
  $\psi(4415)$ & 4.42 \\
  $X(4660)$ & 4.64 \\
  \bottomrule
\end{tabular}
\end{table}
\begin{table}
\caption{Particle Data Group masses of vector bottomonium resonances~\cite{Olive:2016xmw}.}
\label{bottomoniumTable}
\centering
\begin{tabular}{lS}
  \addlinespace
  \toprule
  Name &  \multicolumn{1}{c}{Mass (\si{GeV})} \\
  \midrule
  $\Upsilon(1S)$ & 9.46 \\
  \addlinespace
  $\Upsilon(2S)$ & 10.02 \\
  \addlinespace
  $\Upsilon(3S)$ &  10.34 \\
  $\Upsilon(4S)$ &  10.58 \\
  \addlinespace
  $\Upsilon(10860)$ & 10.89 \\
  $\Upsilon(11020)$ & 10.99 \\
  \bottomrule
\end{tabular}
\end{table}

\begin{table}
\centering
\caption{A representative collection of hadron models analyzed in the charmonium sector.}
\label{charmoniumModels}
\begin{tabular}{ccccccc}
  \addlinespace
  \toprule
  Model & $m_1$ & $\Gamma_1$ & $m_2$ & $\Gamma_2$ & $m_3$ & $\Gamma_3$\\
   & (\si{GeV}) & (GeV) & (GeV) & (GeV) & (GeV) & (GeV)\\ 
  \midrule
  1 & 3.10 & 0 & - & - & - & - \\
  2 & 3.10 & 0 & 3.73 & 0  & - & - \\
  3 & 3.10 & 0 & 3.73 & 0 & 4.30 & 0 \\
  4 & 3.10 & 0 & 3.73 & 0 & 4.30 & 0.30 \\
  5 & 3.10 & 0 & 3.73 & 0.05 & 4.30 & 0.30 \\
  6 & 3.10 & 0 & - & - & 4.30 & 0 \\
  7 & 3.10 & 0 & - & - & 4.30 & 0.30 \\
  \bottomrule
\end{tabular}
\end{table}

\begin{table}
\centering
\caption{A representative collection of hadron models analyzed in the bottomonium sector.}
\label{bottomoniumModels}
\begin{tabular}{ccccccc}
  \addlinespace
  \toprule
  Model & $m_1$ & $\Gamma_1$ & $m_2$ & $\Gamma_2$ & $m_3$ & $\Gamma_3$\\
   & (GeV) & (GeV) & (GeV) & (GeV) & (GeV) & (GeV)\\ 
  \midrule
  1 & 9.46 & 0 & - & - & - & - \\
  2 & 9.46 & 0 & 10.02 & 0  & - & - \\
  3 & 9.46 & 0 & 10.02 & 0 & 10.47 & 0 \\
  4 & 9.46 & 0 & 10.02 & 0 & 10.47 & 0.22 \\
  \bottomrule
\end{tabular}
\end{table}

Substituting~(\ref{rhoForms}) into~(\ref{subedLSR}) gives
\begin{equation}\label{lsrFinal}
  \lsr\double{\tau}{s_0}=\sum_{i=1}^n \int_{4m^2}^{s_0} e^{-t\tau}
  \rho^{\text{(had)}}_i(t)\dif{t}
\end{equation}
where
\begin{equation}\label{specific_rho}
  \int_{4m^2}^{s_0} e^{-t\tau}\rho^{\text{(had)}}_i(t) dt =
  \begin{cases}
    \xi_i e^{-m_i^2 \tau},\ \Gamma_i=0\\
    \xi_i e^{-m_i^2 \tau}\frac{\sinh\big(m_i \Gamma_i\tau\big)}{m_i\Gamma_i\tau},\ 
      \Gamma_i\neq0
  \end{cases}.
\end{equation}
As a specific example, consider a $\rho^{\text{(had)}}$ that has three resonances
with masses $\{m_1,\,m_2,\,m_3\}$.  If the first two resonances are narrow
(i.e., $\Gamma_1=\Gamma_2=0$)
and the third has $\Gamma_3\neq0$, then
\begin{equation}\label{specific_rho2}
  \rho^{\text{(had)}}(t) = \xi_1 \delta(t-m_1^2)+\xi_2 \delta(t-m_2^2)
  +\frac{\xi_3}{2m_3 \Gamma_3}
  \theta\big(t-m_3(m_3-\Gamma_3)\big)\theta\big(m_3(m_3+\Gamma_3)-t\big)
\end{equation}
and
\begin{equation}
  \lsr\double{\tau}{s_0} = 
    \xi_1 e^{-m_1^2 \tau}
    + \xi_2 e^{-m_2^2 \tau}
    + \xi_3 e^{-m_3^2 \tau}\frac{\sinh\big(m_3\Gamma_3\tau\big)}{m_3\Gamma_3\tau}.
\end{equation}

For particular choices of $\{m_i\}_{i=1}^n$ and $\{\Gamma_i\}_{i=1}^n$, the
quantities $\{\xi_i\}_{i=1}^n$ and $s_0$ are extracted as best-fit parameters
to~(\ref{lsrFinal}).
More precisely, we partition the Borel interval
$\double{\tau_{\text{min}}}{\tau_{\text{max}}}$ into $N=20$ equal length
subintervals with $\{\tau_j\}_{j=0}^N$ and define
\begin{equation}\label{chiSquaredGeneral}
  \chi^2(\xi_1,\ldots,\xi_n,\,s_0)=\sum_{j=0}^N
  \Bigg(
    \lsr\double{\tau_j}{s_0}-
    \sum_{i=1}^n \int_{4m^2}^{s_0} e^{-t\tau_j}
    \rho^{\text{(had)}}_i(t)\dif{t}
  \Bigg)^2.
\end{equation}
With the specific $\rho^{\text{(had)}}(t)$ given in~(\ref{specific_rho2}),
for example, eqn.~(\ref{chiSquaredGeneral}) becomes
\begin{equation}\label{chiSquaredDefn}
  \chi^2(\xi_1,\,\xi_2,\,\xi_3,\,s_0)=\sum_{j=0}^N
  \Bigg(
    \lsr\double{\tau_j}{s_0}-
    \xi_1 e^{-m_1^2 \tau_j}
    - \xi_2 e^{-m_2^2 \tau_j}
    - \xi_3 e^{-m_3^2 \tau_j}\frac{\sinh\big(m_3\Gamma_3\tau_j\big)}{m_3\Gamma_3\tau_j}
  \Bigg)^2.
\end{equation}
Minimizing~(\ref{chiSquaredGeneral}) gives predictions for $\{\xi_i\}_{i=1}^n$
and $s_0$ corresponding to the best fit agreement between QCD and the hadronic model
in question.
For the models defined in Tables~\ref{charmoniumModels} and~\ref{bottomoniumModels}, 
our results are shown in Tables~\ref{charmoniumResults} 
and~\ref{bottomoniumResults} respectively.
Rather than present each $\xi_i$, we instead present 
$\zeta$ and $\frac{\xi_i}{\zeta}$ where
\begin{equation}\label{zeta}
  \zeta=\sum_{i=1}^n |\xi_i|.
\end{equation}
The errors included are associated with
the strong coupling reference values~(\ref{alphatau})--(\ref{alphaZ}),
the quark mass parameters~(\ref{charmMass})--(\ref{bottomMass}),
the condensates~(\ref{glueFourDValue})--(\ref{quarkThreeDValue}),
and an allowed $\pm0.1$~GeV variability in the renormalization 
scale~\cite{Narison:2014ska}.
We also allow for the end points of the Borel interval to vary by half the value of 
$\tau_{\text{min}}$, i.e., 
$0.05\ \text{GeV}^{-2}$ in the charmonium sector and
$0.005\ \text{GeV}^{-2}$ in the bottomonium sector. 
We don't vary $\kappa$ from~(\ref{sixDQcond}) as the numerical contribution 
to the LSR~(\ref{finalTheoryLSR}) stemming from the 6d quark condensate diagram is negligible. 
Our results are most sensitive to varying the quark mass parameters.

\begin{table}
\centering
\caption{Predicted mixing parameters with their theoretical uncertainties and continuum thresholds for hadron models
  defined in Table~\ref{charmoniumModels}.}
\label{charmoniumResults}
\begin{tabular}{cSSSSSS}
\addlinespace
\toprule
  Model & \multicolumn{1}{c}{$s_0$} & {$\chi^2\times10^6$} 
    & \multicolumn{1}{c}{$\zeta$} 
    & {$\frac{\xi_1}{\zeta}$} 
    & \multicolumn{1}{c}{$\frac{\xi_2}{\zeta}$} 
    & \multicolumn{1}{c}{$\frac{\xi_3}{\zeta}$} \\
   & {($\text{GeV}^2$)} & {($\text{GeV}^{12}$)} & {($\text{GeV}^{6}$)} & & &\\
\midrule
  1 & 12.5 & 4.33 & 0.514 \pm 0.021 & 1 & {-}  & \multicolumn{1}{c}{-} \\
  2 & 13.9 & 3.17 & 0.734 \pm 0.040 & 0.726 \pm 0.034 & 0.274 \pm 0.034 & \multicolumn{1}{c}{-} \\
  3 & 24.1 & 0.164 & 2.88 \pm 0.25 & 0.215 \pm 0.012 & -0.022 \pm 0.049 & 0.762 \pm 0.030 \\
  4 & 24.2 & 0.162 & 2.97 \pm 0.26 & 0.210 \pm 0.012 & -0.032 \pm 0.048 & 0.758 \pm 0.025 \\
  5 & 24.2 & 0.162 & 2.97 \pm 0.26 & 0.210 \pm 0.012 & -0.032 \pm 0.048 & 0.758 \pm 0.025 \\
  6 & 23.7 & 0.184 & 2.68 \pm 0.25 & 0.228 \pm 0.019 & {-}              & 0.772 \pm 0.019 \\
  7 & 23.6 & 0.204 & 2.66 \pm 0.25 & 0.228 \pm 0.020 & {-}              & 0.772 \pm 0.019 \\
\bottomrule
\end{tabular}
\end{table}

\begin{table}
\centering
\caption{Predicted mixing parameters with their theoretical uncertainties and continuum thresholds for hadron models defined in Table~\ref{bottomoniumModels}.}
\label{bottomoniumResults}
\begin{tabular}{cSSSSSS}
\addlinespace
\toprule
  Model & {$s_0$} & {$\chi^2 \times 10^4$} 
  & {$\zeta$} 
  & {$\frac{\xi_1}{\zeta}$} 
  & {$\frac{\xi_2}{\zeta}$}
  & {$\frac{\xi_3}{\zeta}$} \\
   & {($\text{GeV}^2$)} & {($\text{GeV}^{12}$)} & {($\text{GeV}^{6}$)} & & &\\
\midrule
  1 & 107 & 42.0 & 140 \pm 3 & 1 & {-} & {-} \\
  2 & 100 & 36.5 & 189 \pm 9 & 0.774 \pm 0.014 & -0.226 \pm 0.014 & {-} \\
  3 & 132 & 0.0860 & 1377 \pm 33 & 0.203 \pm 0.002 & -0.380 \pm 0.003 & 0.418 \pm 0.005 \\
  4 & 132 & 0.0879 & 1375 \pm 32 & 0.203 \pm 0.002 & -0.379 \pm 0.003 & 0.418 \pm 0.005 \\
\bottomrule
\end{tabular}
\end{table}

%%%%%%%%%%%%%%% Discussion %%%%%%%%%%%%%%%
\section{Discussion}\label{V}
As can be seen from Tables~\ref{charmoniumResults} and~\ref{bottomoniumResults},
in both the charmonium and bottomonium sectors, the inclusion of a third heavy
resonance cluster in the analysis significantly improves the fit between
QCD and experiment as measured by~(\ref{chiSquaredGeneral}).
The improvement is particularly dramatic for bottomonium.
It is important to note that these third resonance clusters 
make large contributions to the LSR, i.e., the right-hand side 
of~(\ref{lsrFinal}), despite the fact that high mass states are 
suppressed relative to low mass states due to the exponentially
decaying kernel.
As a quantitative measure of the excited state signal strength, consider
\begin{equation}\label{signalStrength}
  \frac{\int_{4m^2}^{s_0} e^{-t\tau}\rho_3^{(\text{had})}(t)\dif{t}}%
  {\sum_{i=1}^{3}\left|\int_{4m^2}^{s_0} e^{-t\tau}\rho_i^{(\text{had})}(t)\dif{t}\right|},
\end{equation}
the ratio of the third resonance's net contribution to the LSR to the 
sum (of the magnitudes) of the contributions made by all three
resonances.
In the charmonium sector,
evaluating~(\ref{signalStrength}) for model~3 from Table~\ref{charmoniumResults}
gives~0.43.
In the bottomonium sector, 
evaluating~(\ref{signalStrength}) for model~3 from Table~\ref{bottomoniumResults}
gives~0.35. Thus the signal strength of the excited state is significant, 
as expected by its clear effect of reducing the $\chi^2$-values in Tables~\ref{charmoniumResults} and~\ref{bottomoniumResults}. 

Including one or more resonance widths in the analysis has almost no impact on the 
quality of fit between QCD and experiment as can be seen from the value of the 
minimized $\chi^2$ of model~3--5 in Table~\ref{charmoniumResults}
and models~3--4 in Table~\ref{bottomoniumResults}.
This is unsurprising given the general insensitivity of LSRs to resonance width.

In both charmonium and bottomonium sectors, including a fourth resonance
or resonance cluster in $\rho^{(\text{had})}$ leads to a $\chi^2$
that minimizes at $s_0\approx m_4^2$, i.e., the heaviest resonance essentially
merges with the continuum, contrary to the initial assumption articulated
in~(\ref{ResCont}) that there is a separation between 
resonance physics and the continuum.
Furthermore, as can be seen from both Tables~\ref{charmoniumResults}
and~\ref{bottomoniumResults}, the two-resonance scenario model~2 
also suffers from this
problem which gives us another reason to disfavour it 
compared to the three-resonance models.

Focusing on the three-resonance models in the charmonium 
sector (model~3--5 in Table~\ref{charmoniumResults}), 
we find  a nonzero mixing parameter for the $J/\psi$; 
essentially no evidence for mixing in the $\psi(2S),\,\psi(3770)$ 
resonance cluster; and
a large mixing parameter corresponding to a resonance (or resonance
cluster) of mass (or average mass) $4.3$~GeV.
We investigated the effect of varying the mass of
the third resonance, $m_3$, from 4.0~GeV--4.6~GeV.
We found that the minimum value of the $\chi^2$ was indeed lowest for
$m_3=4.3$~GeV, about one-third the value for either $m_3=4.0$~GeV
or $m_3=4.6$~GeV.

Given the lack of evidence for meson-hybrid mixing in the $\psi(2S),\,\psi(3770)$ 
resonance cluster, it is reasonable to exclude it from $\rho^{(\text{had})}$. 
As can be seen from models~6--7 in Table~\ref{charmoniumResults}, doing so
has a small effect on the fitted values of $\xi_1,\ \xi_3,$ and $s_0$ 
as well as the minimum value of the $\chi^2$.

Focusing on the three-resonance models in the bottomonium 
sector (models~3--4 in Table~\ref{bottomoniumResults}), 
we find a nonzero mixing parameter for all three resonances, i.e., the
$\Upsilon(1S)$, the $\Upsilon(2S)$, and the $\Upsilon(3S),\,\Upsilon(4S)$ 
resonance cluster, indicating that all have $q\overline{q}$-meson and hybrid components.

In summary, the best agreement between our QCD predictions and experiment is achieved with
three-resonance models in both the charmonium and the bottomonium sectors
although, in the charmonium sector, omitting the second heaviest resonance cluster
has minimal effect on the results.
In fact, $q\overline{q}$-meson-hybrid mixing in the charmonium sector is well-described by a
two resonance model consisting of the $J/\psi$ and a second state with mass 
4.3~GeV.
It has been hypothesized that the $X(4260)$ might be a resonance with a significant hybrid 
component~\cite{ClosePage1995a,Kou:2005gt,Zhu:2005hp}.
Our results are certainly consistent with this idea.
In the bottomonium sector, our results indicate that there is nonzero $q\overline{q}$-meson-hybrid
mixing in the $\Upsilon(1S)$, the $\Upsilon(2S)$, 
and in the $\Upsilon(3S),\,\Upsilon(4S)$ pair.

%%% Acknowledgements and Bibliography %%%
\clearpage
\section*{Acknowledgements}
We are grateful for financial support from the National Sciences and 
Engineering Research Council of Canada (NSERC).

\bibliographystyle{h-physrev}
\bibliography{research}

\begin{thebibliography}{10}

\bibitem{Meyer:2015eta}
C.~A. Meyer and E.~S. Swanson,
\newblock Prog. Part. Nucl. Phys. {\bf 82}, 21 (2015), 1502.07276.

\bibitem{Brambilla:2010cs}
N.~Brambilla {\em et~al.},
\newblock Eur. Phys. J. {\bf C71}, 1534 (2011), 1010.5827.

\bibitem{Eidelman:2012vu}
S.~Eidelman, B.~K. Heltsley, J.~J. Hernandez-Rey, S.~Navas, and C.~Patrignani,
\newblock (2012), 1205.4189.

\bibitem{BESIII:2016adj}
BESIII, M.~Ablikim {\em et~al.},
\newblock Phys. Rev. Lett. {\bf 118}, 092002 (2017), 1610.07044.

\bibitem{BarnesCloseSwanson1995}
T.~Barnes, F.~E. Close, and E.~S. Swanson,
\newblock Phys. Rev. {\bf D52}, 5242 (1995).

\bibitem{Olive:2016xmw}
Particle Data Group, C.~Patrignani {\em et~al.},
\newblock Chin. Phys. {\bf C40}, 100001 (2016).

\bibitem{Shifman:1978bx}
M.~A. Shifman, A.~I. Vainshtein, and V.~I. Zakharov,
\newblock Nucl. Phys. {\bf B147}, 385 (1979).

\bibitem{Shifman:1978by}
M.~A. Shifman, A.~I. Vainshtein, and V.~I. Zakharov,
\newblock Nucl. Phys. {\bf B147}, 448 (1979).

\bibitem{Reinders:1984sr}
L.~J. Reinders, H.~Rubinstein, and S.~Yazaki,
\newblock Phys. Rept. {\bf 127}, 1 (1985).

\bibitem{narisonbook:2004}
S.~Narison,
\newblock {\em QCD as a Theory of Hadrons}, Cambridge Monographs on Particle
  Physics, Nuclear Physics and Cosmology Vol.~17 (Cambridge University Press,
  New York, 2004).

\bibitem{Wilson:1969zs}
K.~G. Wilson,
\newblock Phys. Rev. {\bf 179}, 1499 (1969).

\bibitem{Narison:1984bv}
S.~Narison, N.~Pak, and N.~Paver,
\newblock Phys. Lett. {\bf 147B}, 162 (1984),
\newblock [,77(1984)].

\bibitem{Harnett:2008cw}
D.~Harnett, R.~T. Kleiv, K.~Moats, and T.~G. Steele,
\newblock Nucl. Phys. {\bf A850}, 110 (2011), 0804.2195.

\bibitem{Chen:2013pya}
W.~Chen, H.-y. Jin, R.~T. Kleiv, T.~G. Steele, M.~Wang, and Q.~Xu,
\newblock Phys. Rev. {\bf D88}, 045027 (2013), 1305.0244.

\bibitem{ho:2017}
J.~Ho, D.~Harnett, and T.~G. Steele,
\newblock JHEP\!\! , In Press (2017).

\bibitem{GovaertsReindersWeyers1985}
J.~Govaerts, L.~J. Reinders, and J.~Weyers,
\newblock Nucl. Phys. {\bf B262}, 575 (1985).

\bibitem{PascualTarrach1984}
P.~Pascual and R.~Tarrach,
\newblock {\em {QCD}: Renormalization for the Practitioner} (Springer, 1984).

\bibitem{BaganAhmadyEliasEtAl1994}
E.~Bag\'{a}n, M.~R. Ahmady, V.~Elias, and T.~G. Steele,
\newblock Z. Phys. {\bf C61}, 157 (1994).

\bibitem{AkyeampongDelbourgo1973}
D.~A. Akyeampong and R.~Delbourgo,
\newblock Nuovo Cim. {\bf 17}, 578 (1973).

\bibitem{MertigScharf1998}
R.~Mertig and R.~Scharf,
\newblock Comput. Phys. Commun. {\bf 111}, 265 (1998).

\bibitem{Tarasov1996}
O.~V. Tarasov,
\newblock Phys. Rev. {\bf D54}, 6479 (1996).

\bibitem{Tarasov1997}
O.~V. Tarasov,
\newblock Nucl. Phys. {\bf B502}, 455 (1997).

\bibitem{BoosDavydychev1991}
E.~E. Boos and A.~I. Davydychev,
\newblock Theor. Math. Phys. {\bf 89}, 1052 (1991).

\bibitem{BroadhurstFleischerTarasov1993}
D.~J. Broadhurst, J.~Fleischer, and O.~V. Tarasov,
\newblock Z. Phys. {\bf C60}, 287 (1993).

\bibitem{AbramowitzStegun1965}
M.~Abramowitz and I.~A. Stegun,
\newblock {\em Handbook of Mathematical Functions} (Dover Publications, 1965).

\bibitem{Narison:1981ts}
S.~Narison and E.~de~Rafael,
\newblock Phys. Lett. {\bf B103}, 57 (1981).

\bibitem{Launer:1983ib}
G.~Launer, S.~Narison, and R.~Tarrach,
\newblock Z. Phys. {\bf C26}, 433 (1984).

\bibitem{Narison2010}
S.~Narison,
\newblock Phys. Lett. {\bf B693}, 559 (2010).

\bibitem{ChenKleivSteeleEtAl2013}
W.~Chen, R.~T. Kleiv, T.~G. Steele, B.~Bulthuis, D.~Harnett, J.~Ho,
  T.~Richards, and S.-L. Zhu,
\newblock Journal of High Energy Physics {\bf 1309}, 019 (2013).

\bibitem{BergHarnettKleivEtAl2012}
R.~Berg, D.~Harnett, R.~T. Kleiv, and T.~G. Steele,
\newblock Phys. Rev. {\bf D86}, 034002 (2012).

\bibitem{Harnett:2012gs}
D.~Harnett, R.~T. Kleiv, T.~G. Steele, and H.-y. Jin,
\newblock J. Phys. {\bf G39}, 125003 (2012), 1206.6776.

\bibitem{Narison:2014ska}
S.~Narison,
\newblock Int. J. Mod. Phys. {\bf A30}, 1550116 (2015), 1404.6642.

\bibitem{ClosePage1995a}
F.~E. Close and P.~R. Page,
\newblock Phys. Rev. {\bf D52}, 1706 (1995).

\bibitem{Kou:2005gt}
E.~Kou and O.~Pene,
\newblock Phys. Lett. {\bf B631}, 164 (2005), hep-ph/0507119.

\bibitem{Zhu:2005hp}
S.-L. Zhu,
\newblock Phys. Lett. {\bf B625}, 212 (2005), hep-ph/0507025.

\end{thebibliography}

\end{document}